# Effect of dipole-dipole charge interactions on dust coagulation

**L S Matthews and T W Hyde**
Center for Astrophysics, Space Physics, and Engineering Research, Baylor University, Waco, TX, USA 76798-7316

E-mail: Lorin_Matthews@baylor.edu

**Abstract.** This study examines the effect that dipole-dipole charge interactions between fractal aggregates have on the growth of dust grains. Aggregates in a plasma or radiative environment will have charge distributed over their extended surface, which leads to a net dipole moment for the charged grains. A self-consistent N-body code is used to model the dynamics of interacting charged aggregates. The aggregates are free to rotate due to collisions and dipole-dipole electrostatic interactions. These rotations are important in determining the growth rate and subsequent geometry (fractal dimension) of the grains. In contrast to previous studies which have only taken charge-dipole interactions into account, like-charged grains are found to coagulate more efficiently than neutral grains due to preferential incorporation of small aggregates into mid-sized aggregate structures. The charged aggregates tend to be more compact than neutral aggregates, characterized by slightly higher fractal dimensions.

## 1. Introduction

Dust, a ubiquitous component of the universe consisting of particles ranging in size from nanometers to millimeters, can become charged in astrophysical environments and laboratory applications. Charged dust levitates above the surface of the moon (Horányi 1996), is responsible for the appearance of the "spokes" in Saturn's rings (Goertz and Morfill 1983), and is an undesirable by-product in industrial plasma processing systems and fusion reactors (Selwyn et al 1990, Winter 1998).

The growth of dust aggregates through collisions between the grains constitutes a fundamental process in dusty plasma physics. Charges on the grains play an important role in determining both their growth rate and the maximum size the resulting aggregates can attain through collisional growth: local plasma conditions determine whether grains acquire an equilibrium charge of the same or opposite sign. Oppositely charged grains can undergo rapid coagulation, or charge gelation, in which a majority of the particles are collected into a single large aggregate. This phenomenon is commonly ascribed to the charge-dipole interactions between the grains and has been observed both experimentally (Konopka *et al* 2005) and numerically (Matthews and Hyde 2004). In low temperature plasma environments, grain charging occurs due to the collection of ions and electrons on the surface of the grain. In this case, most of the grains are charged to a negative potential due to the greater electron mobility. Thus, the Coulomb interaction between grains is repulsive, inhibiting grain growth. Interestingly, the coagulation rate for such like-charged grains may be enhanced if charge-dipole interactions, which can be created by an anisotropic plasma flow (Lapenta 1998) or induced dipoles (Simpson *et al* 1979), are included. The physical geometry of the aggregates produced by collisions can also lead to a net dipole moment on the grain, which can also have a subsequent effect on the coagulation process (Matthews and Hyde 2008).

Dust particles colliding at low velocities tend to stick at the point of contact (Dominik and Tielens 1997), producing fluffy, fractal aggregates much like the fluffy dust bunnies found under a bed. Since these dust aggregates in general have an irregular structure, the charge is not symmetrically distributed around the aggregate's center of mass making it necessary to model the grain's resulting electrostatic potential using a multipole expansion. Including dipole-dipole interactions can cause interacting aggregates to rotate as they approach one another, collide, and stick, influencing the overall orientation and structure of the new aggregate. This can result in a more compact object, which has a smaller cross-sectional area for future collisions, or a more open, extended object, having a greater cross-sectional area.

This study extends previous analyses examining the coagulation of charged fractal aggregates by including the effect of dipole-dipole interactions between aggregates. The manner in which the charge on a fractal aggregate is determined is described in Section 2. The numerical methods used to build the aggregates and model the coagulation process are described in Section 3. In Section 4, the evolution of dust populations are examined to determine the effect of the dipole-dipole interactions. Conclusions are presented in Section 5.

## 2. Charging of Fractal Aggregates

Calculating the charge on a single spherical dust grain immersed in a plasma environment is relatively simple under the proper set of assumptions (Allen 1992, Laframboise and Parker 1973, Lampe *et al* 2001); calculating the charge on a non-spherical object under the same conditions is much more difficult. Determining the charge on an extended, irregular structure such as a fractal aggregate is harder yet and can only be accomplished numerically. The charge on such an aggregate is necessarily distributed over its surface, which can lead to an effective dipole moment for the grain. In this study, all aggregates considered are assembled from an irregular collection of spherical monomers (figure 1).

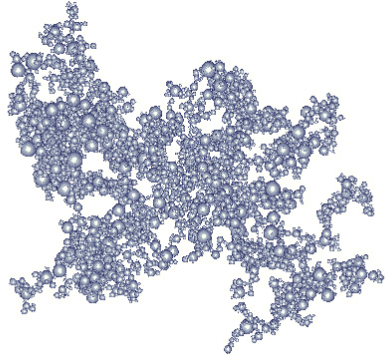

Figure 1. Representative aggregate built from an initial population having a power law size distribution. The number of monomers in the aggregate is $N = 11037$, and the maximum radial extent from the aggregate's center of mass is $r_{max} = 579$ μm.

The surface potential of a single, small isolated grain is most often calculated using OML theory. This theory is based on the assumption that energy and momentum are conserved for impinging current species and that ions and electrons which have encountered potential barriers (such as those where their trajectories intersect another grain) have been removed from the background Maxwellian distribution (Kennedy and Allen 2003, Laframboise and Parker 1973, Lampe *et al* 2001). Non-spherical aggregates must be treated in a different manner, since the assumption that the motion of electrons and ions takes place in a central field is no longer valid. As a first approximation, a line-of-sight (LOS) approach is used to calculate the total flux to each monomer within an aggregate. The equilibrium charge on a grain can be determined by imposing the condition that the sum of all currents to the grain must be zero. The current density $J$ of species $\alpha$ (usually electrons or ions) to any point on a grain is given by

$$J_\alpha = n_{\alpha\infty} q_\alpha \int_{v_{min,\alpha}}^{\infty} f_\alpha v_\alpha \cos(\theta) d^3 \vec{v}_\alpha \qquad (1)$$

where $n_{\alpha\infty}$ is the number density outside the grain's potential well, $q_\alpha$ is the charge on the plasma species, $f_\alpha$ is the distribution function, $\theta$ is the angle between the radial position $r$ and velocity $v$, and the integration is carried out over the velocity space $d^3\vec{v}_\alpha$ of all orbits intersecting the surface for the first time (Allen 1992). The lower limit of integration is defined as the minimum velocity of a charged particle, incident from infinity, which will allow it to reach the surface of the grain. By making the substitution $d^3\vec{v}_\alpha = v_\alpha^2 dv_\alpha d\Omega$, the integrals over the speed and angles may be done independently, resulting in

$$J_\alpha = n_{\alpha\infty} q_\alpha \int_{v_{\min,\alpha}}^{\infty} f_\alpha v_\alpha^3 dv_\alpha \int \cos(\theta) d\Omega \,. \tag{2}$$

By specifying a plasma distribution function appropriate for the given environment, the integration over velocity can be readily carried out. The calculation of charge on a fractal aggregate is then reduced to determining the solid angle, $d\Omega$, for unobstructed orbits to each monomer within the aggregate. A LOS approximation is used to determine which orbits to exclude from the limits of integration (Matthews and Hyde 2008). It is assumed that electrons or ions move in a straight line from infinity and that the flux to any point on the surface of a monomer from a direction whose line of sight is blocked by a grain, including itself, is excluded from the integration in Eq. (2) (see figure 2). After calculating the LOS factor, $d\Omega$, separately for each constituent monomer, the integration in Eq. (2) is then carried out over the entire surface of the aggregate. One should note that this approximation becomes less accurate as a grain becomes more highly charged, since ion orbits can then have significant curvature in the vicinity of the charged grain.

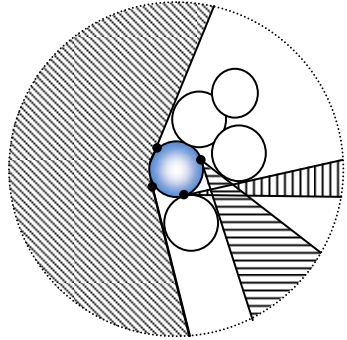

Figure 2. The LOS factor $d\Omega$ is determined for each monomer by finding the unblocked trajectories of ions and electrons incident from infinity to given points on the monomer. The shaded regions on this 2D cross section indicate the contributions to the LOS factor from four points on a monomer's surface.

Assuming that the aggregate consists of a dielectric material, the impinging charged species will remain near the point of impact. This allows the monopole and dipole contributions to the potential to be calculated for individual monomers, with the sum of these determining the monopole and dipole potential of the entire aggregate. This new potential is then used to recalculate the charging currents to the grains and the process repeated until the grains attain an equilibrium potential.

The OML_LOS approximation was used to calculate the charge and dipole moment of a large number of fractal aggregates which were assembled as described in Section 3.1. For illustrative purposes and to compare results with previous studies, the grains were charged assuming a plasma temperatures $T_e = T_i$ leading to a potential of -1 V on a spherical dust grain. The equilibrium charge on an aggregate $Q$, was then compared to the charge the aggregate would have if simple charge conservation was assumed during the coagulation process, $Q_o$. The results indicate that $Q$ can be well-approximated as a function of $Q_o$ and $N$, as indicated by the linear fit on a log-log plot as shown in figure 3a (Matthews and Hyde 2008). This

provides a heuristic charging scheme which can be used to quickly calculate the charge on a fractal aggregate

$$Q = AQ_oN^B \tag{3}$$

where $A$ and $B$ are constants determined by the equation for the linear fit.

A similar analysis can be used to predict the dipole moment of a fractal aggregate. In this case, the magnitude of the dipole moment, $|\mathbf{p}|$, calculated by OML_LOS is compared to the dipole moment found assuming charge conservation, $|\mathbf{p}_o|$, where the charge is arranged over the aggregate such that the $i^{th}$ monomer carries a charge

$$q_i = \frac{Qd_i}{\sum d_i} \tag{4}$$

where $d_i$ is the distance of the $i^{th}$ monomer from the center of mass. In this case the linear fit shows a large amount of scatter since the dipole moment depends on the geometry of a given aggregate as well as $N$. Representative results are shown in figure 3b. Deviations from the fit line are generally within an order of magnitude, allowing an approximate dipole moment to be quickly calculated for a given aggregate during coagulation simulations.

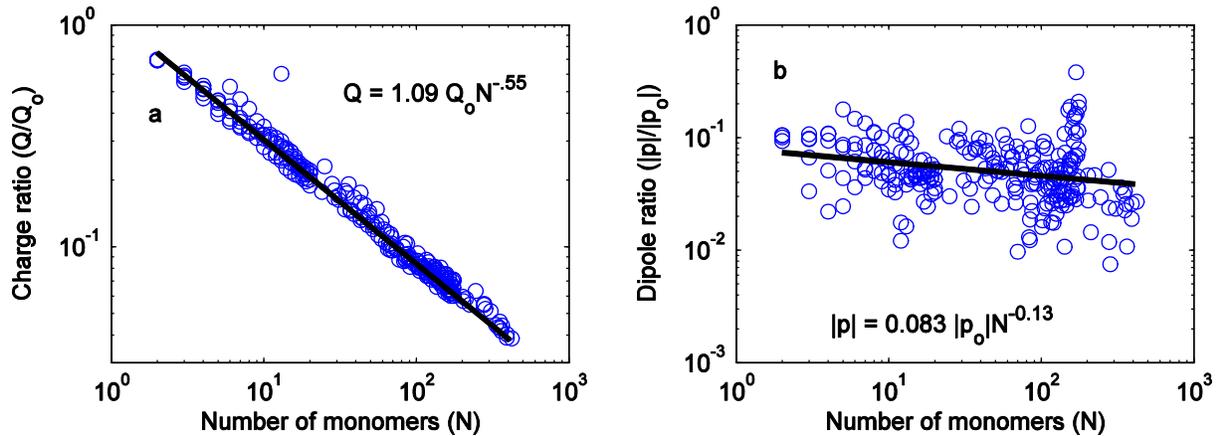

Figure 3. Charges (a) and dipole moments (b) calculated by OML_LOS. The open circles indicate the ratio of the charge and dipole moment calculated using OML_LOS to that calculated using charge conservation during collisions. The trend lines show the linear best fit. The charge scales as $Q \sim N^{-0.55}$, while the dipole moment is dependent on $N$ to a lesser degree.

## 3. Coagulation Models

Several prior studies have examined the growth rate and fractal dimension of dust grains interacting through Brownian motion (Simpson 1978, Meakin 1984, Lapenta 1998, Horányi and Goertz 1990, Ivlev *et al* 2002). Such analytical approaches can provide expected growth rates and particle structures; however, in these models collisions between particles must occur according to a specific set of rules designed to simplify the physics involved. As a result, self-consistent N-body simulations are essential for properly modeling the dynamics of the interacting grains while also resolving collisions. N-body simulations also have the advantage of being able to resolve spatial fluctuations within the dust population (Kempf *et al* 1999) while accurately modeling the three-dimensional geometry of colliding aggregates (Richardson 1995, Matthews and Hyde 2004).

This study uses an N-body code (Matthews and Hyde 2004) to build on previous work analyzing the coagulation of charged aggregates assembled from spherical monomers. Calculation of both coagulation and charging are computationally intensive to model for fractal aggregates, so two methods for reducing the CPU time have been developed. The time required for a grain's charge to reach equilibrium is most strongly dependent on the plasma electron and ion frequencies, which are usually much greater than the collision frequency. Thus the heuristic charging algorithm described in Section 2 can be employed to determine the new charge on the aggregate at the time of collision, eliminating the CPU intensive process of performing a detailed charging calculation from the coagulation model.

The most CPU intensive portion of an N-body simulation occurs during its initial development period when the majority of the particles are monomers. To build realistic aggregates with hundreds or thousands of monomers requires starting with tens of thousands of particles. To minimize this problem, an Aggregate Builder code has been developed to quickly build large aggregates from monomers and smaller aggregates by modeling potential collisions between particle pairs (Matthews *et al* 2007). Smaller aggregates are assembled through the addition of single monomers with the resulting aggregates then employed in subsequent collisions to build medium aggregates. Finally, large aggregates are assembled through collisions between both small and medium-sized aggregates. At each stage, data on all aggregates are saved to a library to be used as input in more complex N-body simulations.

For this study, N-body simulations were run employing initial populations of aggregates generated by Aggregate Builder (Matthews *et al* 2007). Dust grains in astrophysical environments are observed to have a power law spectrum in radius $n(a)da = a^{\gamma} da$, where $n(a) da$ represents the number of particles with radii in the range $(a, a+da)$. Numerical and experimental studies commonly use grains with radii on the order of 1 μm as representative of dust found in protoplanetary disks (Blum and Wurm, 2008). These studies also often employ monomers of equal size, with the justification that Brownian motion tends to form clusters of similar size (Kempf, Pfalzner, and Henning 1999). Therefore, two different initial monomer populations were used. Population I consisted of a power law distribution with grains ranging in size from 0.5-10 μm with $\gamma = -1.8$ (Weidenschilling and Ruzmaikina 1994). Population II consisted of grains having a narrower size distribution with radii ranging linearly from 1 μm to 6 μm. These two distributions yield a mean monomer radius of a few microns (1.8 μm and 3.5 μm, respectively). The aggregates were assembled up to a size of ~2000 monomers. The velocity of the incoming grains was established as the minimum velocity necessary for two 0.5 μm grains charged to -1V potentials to collide when approaching from infinity, $v_{rel} = 0.326$ m/s. The velocity of larger grains was scaled by the square root of the mass, and then given a Gaussian deviation from this mean.

Five datasets were randomly produced from aggregates built with Aggregate Builder for each of the described populations. Aggregates were given random velocities of a few cm/s, which leads to sticking without restructuring (Dominik and Tielens 1997). The overall aggregate distribution was defined using an exponential size distribution in the aggregate radius, $r_{max}$ (defined to be the radius of a sphere with its origin located at the center of mass which just encompasses the aggregate). The aggregate characteristics for the two populations are compared in Tables 1 and 2. The two populations have a minimum radius differing by a factor of two because of the different minimum monomer radius in the two populations. A power law distribution is sensitive to the minimum size limit and relatively independent of the maximum size, thus Population I had a greater percentage of small aggregates with $N \leq 20$. Initial data sets built from Population I monomers had 500 aggregates with an average total of ~10 000 monomers, while those built from Population II had 300 aggregates with an average total of ~29 000 monomers.   On average, the largest aggregate in a Population I dataset had greater radius (and mass) and smaller N than the largest aggregate in a Population II dataset.

Table 1. Comparison of number of monomers $N$ per aggregate.

|  | Minimum | Mean | Median | Maximum | % with $N \leq 20$ |
|---|---|---|---|---|---|
| Population I | 2 | 22.3 ± 4.7 | 11.8 ± 0.4 | 1930 ± 161 | 88.0 ± 2.0 |
| Population II | 2 | 98.5 ± 8.4 | 14.0 ± 0.8 | 2036 ± 113 | 76.8 ± 1.6 |

Table 2. Comparison of aggregate radius $r_{max}$ (in µm).

|  | Minimum | Mean | Median | Maximum |
|---|---|---|---|---|
| Population I | 1.0 | 23.6 ± 0.9 | 21.6 ± 0.4 | 279 ± 13 |
| Population II | 2.0 | 35.1 ± 1.5 | 20.4 ± 0.4 | 258 ± 18 |

For each of the populations, two cases were run: in the first, the aggregates were uncharged while in the second dust grains interacted under the influence of mutual electrostatic forces with dipole-dipole interactions between grains (Matthews *et al* 2007). The typical charge on an aggregate is such that the (monopole) potential calculated at a distance $r_{max}$ is in the range of -0.5 to -1.0 V. The magnitude of the dipole moment is typically five to six orders of magnitude smaller than the magnitude of the charge. Given that the dipole moment is proportional to charge and distance of separation, if one assumes that the dipole charge is of the same order of magnitude as the total charge, the charge separation is on the order of tens of microns, which is comparable to a typical aggregate radius.

The electric field of particle *i*, $\mathbf{E}_i$, is calculated using the monopole and dipole contributions. The net electric field of all the particles then can induce both an acceleration of and a torque on particle *j*. The torque is given by $\mathbf{N}_j = \mathbf{p}_j \times \mathbf{E}$ and the resulting rotational motion of the aggregate is then solved using Euler's equations (Matthews *et al* 2007). Binary collisions were sufficiently resolved that only "true" collisions were detected; in other words, for two aggregates within a distance smaller than the sum of their maximum radial extent, $R < R_i + R_j$, their orientations were updated and a particle-by-particle check performed to determine any monomers that were in contact (Richardson 1995). In all cases, it was assumed colliding aggregates stick at the point of contact. System evolution was tracked until the number of aggregates remaining was approximately 15% of its initial value.

**4. Results**
The evolution of the dust cloud can be characterized through examination of several different parameters: the coagulation rate, the type of collisions driving the coagulation process, and the fractal dimension of the aggregates.

*4.1 Coagulation Rate*
The coagulation rate is determined by tracking the number of aggregates in the system as a function of time. This data is plotted for the two populations in figure 4. Interestingly, the total number of aggregates decreases more rapidly for charged aggregates, indicating that charged aggregates coagulate more rapidly than do neutral aggregates. This seems unusual given that the coagulation of charged aggregates initially lags behind that of the neutral aggregates, with the number of charged aggregates being given by a bell-shaped curve. The tail end of the bell curve is described by an exponential decrease in time $N \propto e^{-t}$, as indicated in the plots. In contrast, the neutral aggregate population decreases as $N \propto t^{-2}$ throughout the simulation. This difference in the coagulation rate indicates that different processes are driving the evolution of the dust population.

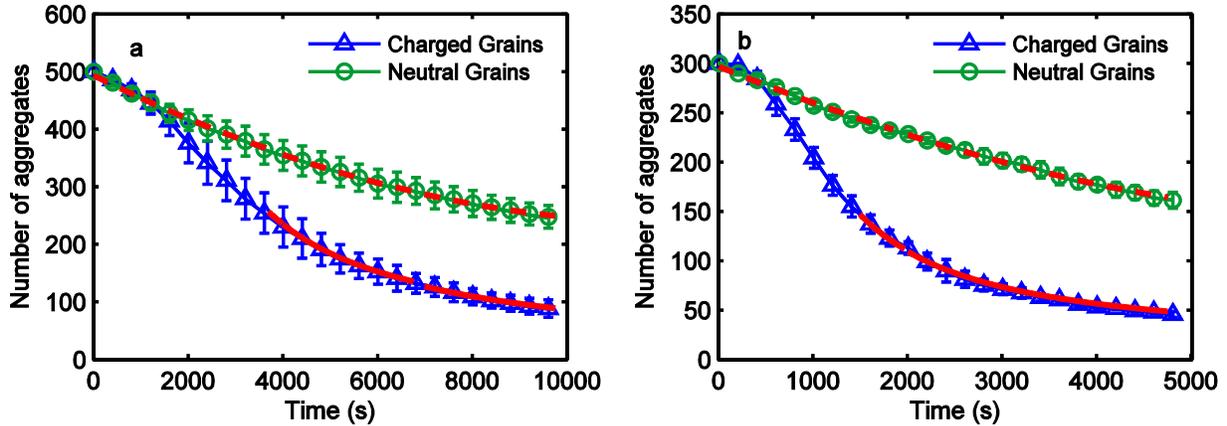

Figure 4. Number of aggregates vs. time. (a) Population I (as defined in text). (b) Population II. While the number of aggregates in the neutral populations decreases as $N \sim t^{-2}$ over the entire simulation (dotted red line), the charged aggregates' rate of increase transitions from a bell-shaped curve to an exponential decrease in time ($N \sim \exp(-t)$), indicated by the solid red line.

*4.2 Collision Types Driving Coagulation*

Examination of the mass and mass distribution of the aggregates provides clues as to the cause behind the difference in coagulation rate between the charged and neutral populations. It has previously been shown that the mean cluster mass in a system undergoing coagulation usually evolves as $m(t) \propto t^z$, where $z$ is defined as the dynamic exponent (van Dongen and Ernst 1985). The dynamic exponent is related to the types of collisions dominating the coagulation process (e.g. large aggregates colliding with large aggregates as opposed to large aggregates colliding with small aggregates). The evolution of the aggregate mean mass for the two populations given above is shown in figure 5. While for neutral aggregates the growth of the mean mass is linear in time throughout the simulation, $\langle m \rangle \propto t$, the mean mass of the charged aggregates evolves from being quadratic in time, $\langle m \rangle \propto t^2$, to linear growth, $\langle m \rangle \propto t$. This transition occurs at approximately the same time the number of aggregates begins to be well characterized by an exponential decay (figure 4).

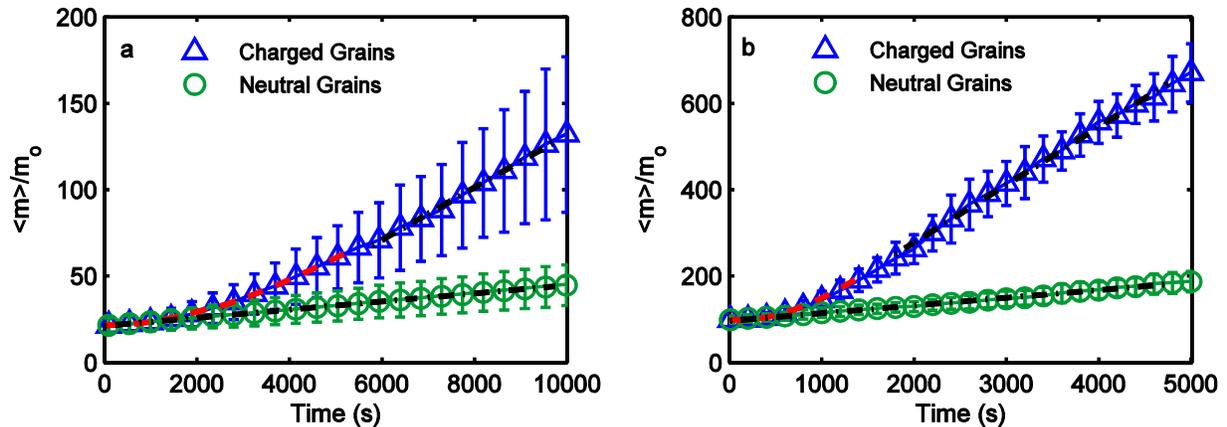

Figure 5. Mean mass of aggregates over time. (a) Population I. (b) Population II. The average mean mass of the charged aggregates initially increases as $t^2$, indicated by the dotted red line. This rate of increase becomes linear in time as the simulation progresses, indicated by the black dotted-dashed line. The rate of increase for neutral aggregates remains linear throughout the simulation.

The types of collisions occurring within the two populations can be inferred by examining the evolution of the mass of the largest aggregate in the system (figure 6) and the evolution of the overall mass distribution (figures 7 and 8). In figure 6, the growth of the largest aggregate in the system shows a stark contrast between charged and neutral grains. The largest charged aggregate grows very slowly over time at almost a constant rate, indicating it is accumulating new mass in small increments, i.e. only through collisions with the smallest aggregates. In this case, mutual electrostatic repulsions prevent larger aggregates from colliding with each other. Only the smallest aggregates, with their higher relative velocities and lower charges, are able to collide and stick with the largest aggregates. The growth of the largest neutral aggregate is characterized by large jumps, occurring when it collides with another large aggregate. The error bars on the growth of the neutral aggregate are very large, reflecting the random nature of the collisions. (The error bars for both charged and neutral grains are larger for Population I particles since the variation in mass of aggregates between initial datasets is much greater for the power-law size distribution.)

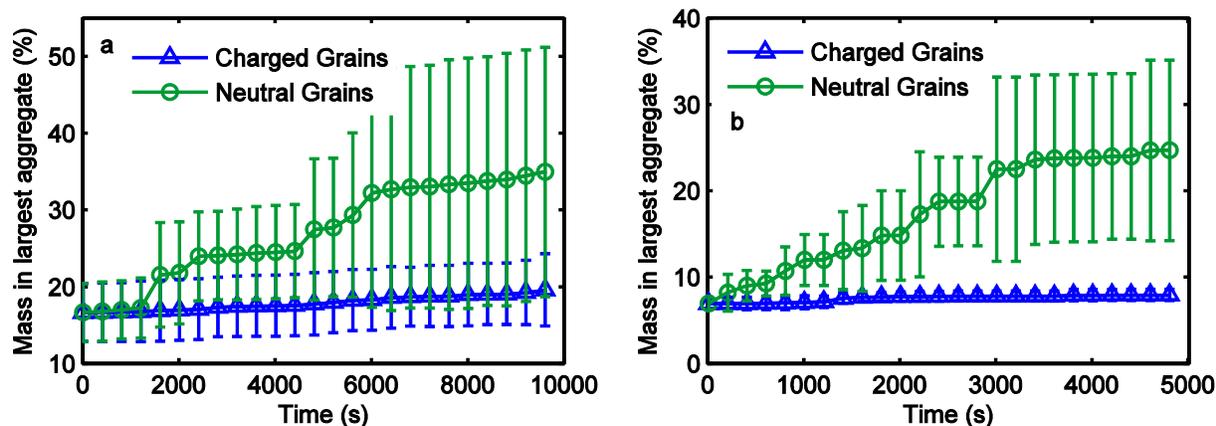

Figure 6. Largest aggregate vs. time. (a) Population I. (b) Population II. The data points indicate the average values for the five data sets, with error bars giving the standard deviation for the data. As shown, Population I has much larger error bars due to the larger mass differences among the constituent monomers. For neutral populations, the largest aggregates grew by collisions with other large aggregates as evidenced by steep jumps in the aggregate mass. The largest charged grains showed slow growth, indicating that the largest grains were only colliding with the smallest grains.

Further differences in these collision mechanisms can be seen in the mass fraction distribution for each population. The grain size evolution is shown in figures 7 and 8 where the mass fraction per logarithmic interval, $M_i = \sum_j m_{ij}/m_{tot}$, is plotted vs. binned aggregate mass for several equally spaced time intervals. Here $i$ denotes the $i^{th}$ logarithmic bin, $j$ is the index of each aggregate populating the bin, and $m_{tot}$ is the total mass of all aggregates. The data shown is the average for the five data sets. By $t = 3300$ s for Population I and $t = 1670$ s for Population II (roughly one-third of the way through the total simulation time), the differential mass distribution has shifted to the right (larger masses) for charged aggregates with masses $\leq 200\ m_o$ (where $m_o$ is the average monomer mass), indicating that these smallest-mass aggregates now constitute a much smaller percentage of or have been completely removed from the total population. As mentioned previously, these times also correspond to the change in the nature of the aggregate number decay rate in the population as seen in figure 4. During the same time interval, very little change occurs in the mass fraction distribution for high-mass aggregates in the charged populations, further evidence that the largest aggregates are not colliding with each other but only growing through collisions with smaller aggregates. The total population tends to evolve towards more mid-sized aggregates dominating the mass spectrum. For the neutral grains, the opposite is true. The distribution shifts very little for the lowest-mass aggregates ($m \leq 30\ m_o$), but both increases in $M_i$ and a shift towards the right can be seen for midsized ($30\ m_o < m \leq 1000\ m_o$) and high-mass ($m > 1000\ m_o$) grains, respectively. Thus, small aggregates are not preferentially removed from the system; rather the largest aggregates tend to sweep up any aggregate within its path.

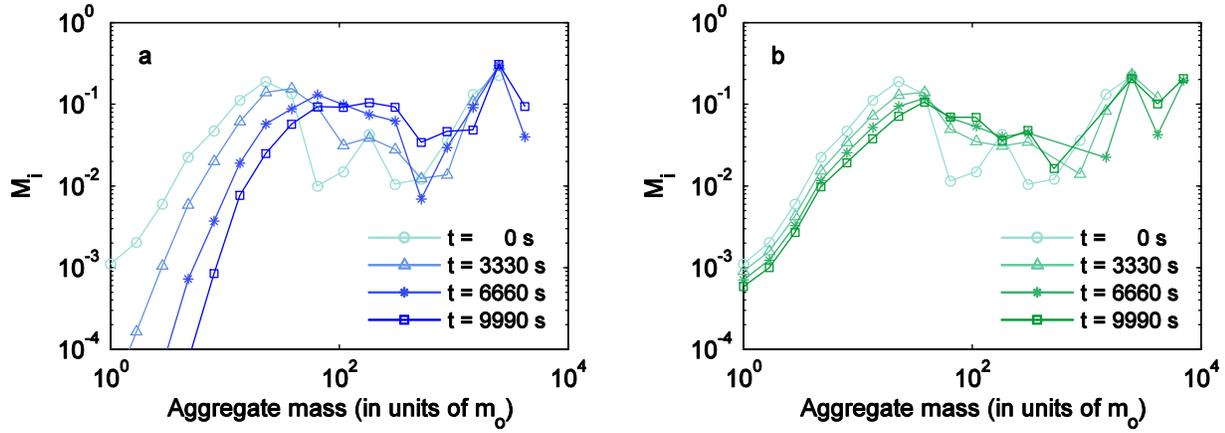

Figure 7. Mass fraction per logarithmic interval -- Population I. (a) The mass fraction curves for the charged aggregates shift rapidly to the right as the smallest aggregates ($m < 30\ m_o$) are depleted. Growth is mainly seen in the mid-sized range ($30\ m_o < m < 300\ m_o$). In contrast, the neutral aggregates (b) have most of their growth occur for the largest aggregates ($m \geq 500\ m_o$) with the appearance of particles in new high-mass bins.

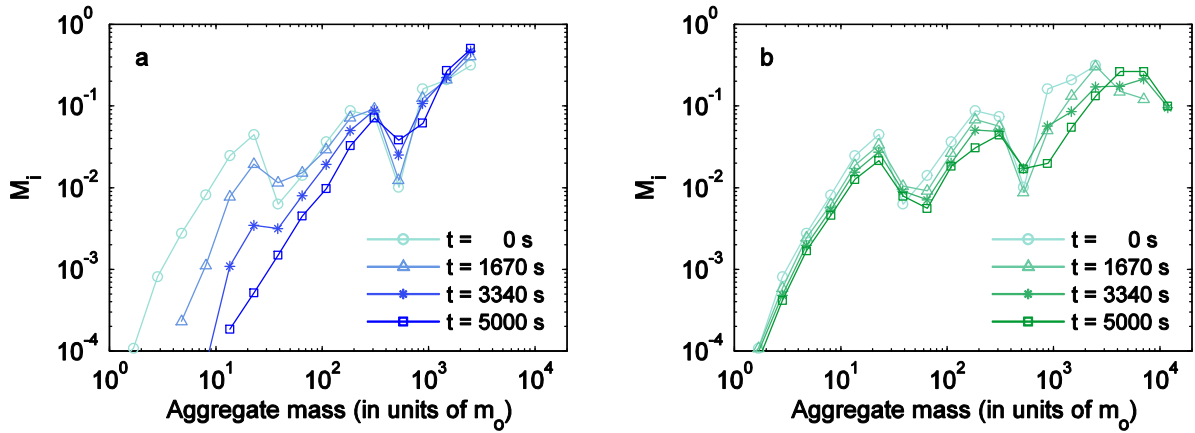

Figure 8. Mass fraction per logarithmic interval -- Population II. (a) The mass fraction curves for the charged aggregates shift rapidly to the right at the smallest aggregates are depleted. In contrast, the neutral aggregates (b) have very small decreases in the mass fraction contained in the smallest mass bins. Most of the growth occurs as largest aggregates collide, shifting the very end of the curve to the right.

### 4.3 Fractal Dimension

Finally, it is instructive to examine the evolution of the aggregates' fractal dimension throughout the coagulation process. In this case, the fractal dimension is calculated by determining the slope of a log-log plot of distance from the center of mass of the aggregate versus the enclosed number of monomers. This method for estimating fractal dimension produces large error bars for aggregates containing only a few monomers; to minimize this problem, the statistics shown in figures 9 and 10 are for aggregates containing ten or more monomers. The probability distribution function for fractal dimensions represents the prevalence a given fractal dimension has within the population, while a plot of the mean aggregate mass versus fractal dimension indicates whether similarly-sized aggregates have similar fractal dimension.

The probability distribution for fractal dimension (figure 9) suggests that for Population I the final mean fractal dimension is essentially the same for both charged and uncharged aggregates, $D_f = 1.89$. However, the probability distribution curve for charged aggregates is much more sharply peaked; in other words, very few aggregates with low fractal dimensions ($D_f < 1.5$) remain at the end of the simulation. The fact that the smallest aggregates ($N < 10$) are not included together with the fact that the smallest charged aggregates are depleted by the end of the simulation indicates the remaining mid-sized aggregates not only have similar masses, but similar fractal dimensions.  Charged and neutral aggregates in Population II show markedly different final probability distributions. It is interesting to note that neutral aggregates in this case have the same mean fractal dimension as do neutral aggregates in Population I. In comparison, charged aggregates have a higher mean fractal dimension, $D_f = 2.07$, and the lack of aggregates having low fractal dimension is even more pronounced.  This indicates that charged aggregates tend to form aggregates of similar fractal dimension by preferentially removing aggregates with small fractal dimension from the population.

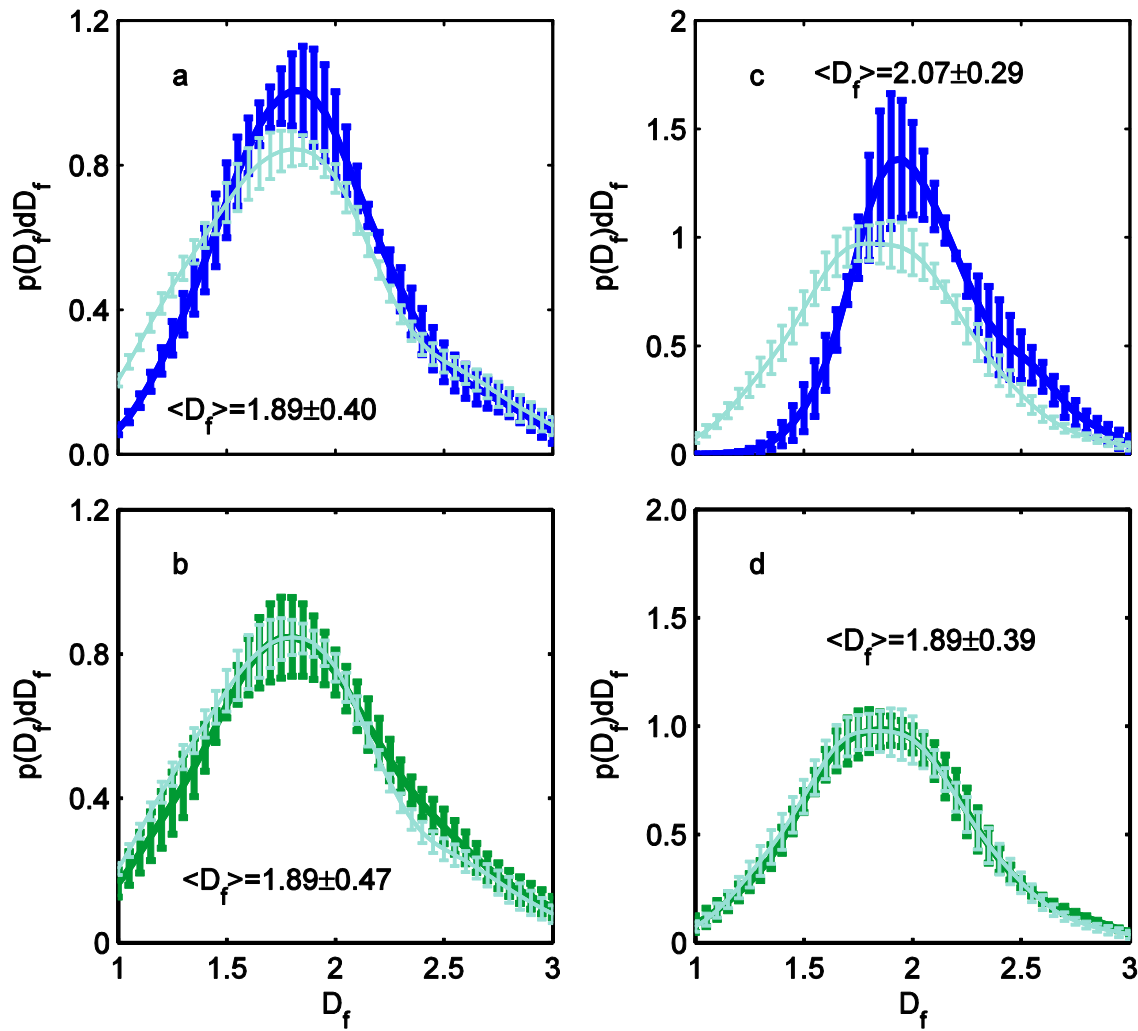

Figure 9. Probability density estimate for fractal dimension. (a, b) Population I charged (blue) and neutral (green) aggregates. The initial distribution in fractal dimension is shown in light blue-green.  While the final mean fractal dimension, indicated on each figure, is essentially the same for both charged and uncharged aggregates, the curve for the charged aggregates is more sharply peaked with very few aggregates with the lowest fractal dimensions ($D_f < 1.5$) remaining at the end of the simulation. (c, d) Population II charged (blue) and neutral (green) aggregates. The initial distribution in fractal dimension is shown in light blue-green. Here charged aggregates have a markedly higher mean fractal dimension and the lack of aggregates with the lowest fractal dimensions is even more pronounced.

The mean aggregate mass vs. fractal dimension for both initial and final distributions is shown in figure 10. In Population I (fig. 10 a, b), charged grains showed significant increases in mass for nearly all fractal dimensions, with the most significant growth seen in the bins for $1.2 < D_f < 1.8$, even though the highest masses tended to occupy the bins $1.8 < D_f < 2.2$. This is in contrast to neutral grains which only showed significant growth in a few of the fractal dimension bins, indicating that only selective aggregates grew significantly larger. In Population II (fig. 10 c, d) charged grains with the smallest fractal dimensions ($D_f < 1.5$) were selectively removed from the population during the aggregation process while the neutral population shows almost no change in these bins. In the charged population, almost all of the bins encompassing $1.7 < D_f < 2.2$ all have aggregates with mass ~1000 $m_o$, with several high mass aggregates having large fractal dimension, $D_f < 2.4$. Again, significant growth in the neutral population occurs only in a few select bins.

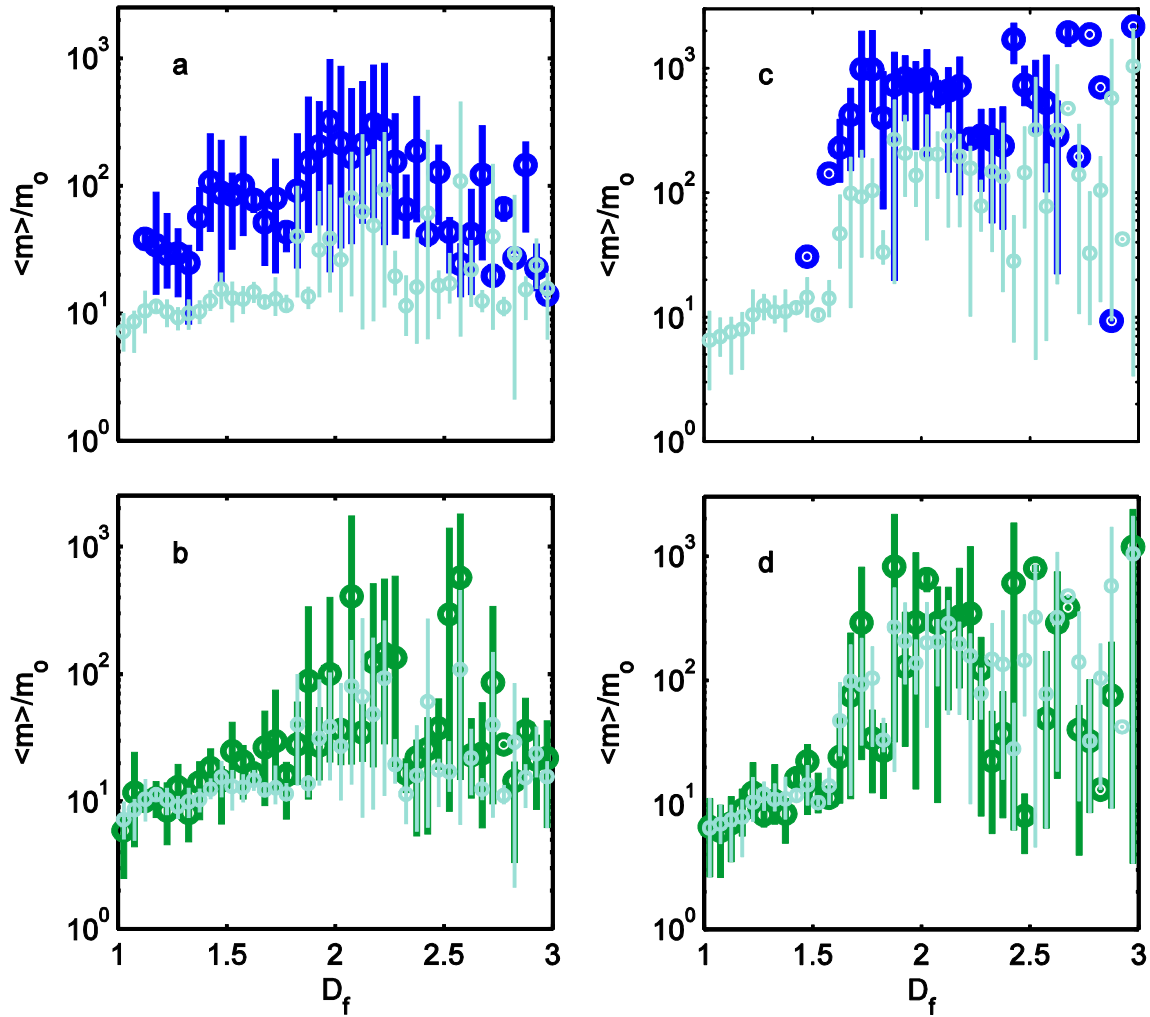

Figure 10. Mean aggregate mass (open circles) in bins for fractal dimension at the beginning (light blue-green) and end of the simulation (blue for charged grains; green for neutral grains). The mean mass

variation of the clusters contributing to each bin is indicated by the vertical lines. (a, b) Population I. The charged grains showed significant increases in mass for nearly all fractal dimensions. This is in contrast to the neutral grains which only showed significant growth in a few of the fractal dimension bins, indicating that only selective aggregates grew significantly larger. (c, d) Population II. The charged grains with the smallest fractal dimensions ($D_f < 1.5$) were selectively removed from the population during the aggregation process. The neutral population shows almost no growth for the smallest fractal dimensions, with significant growth occurring only in select bins.

**5. Conclusion**
A new model of aggregation of charged micron-sized grains has been presented which includes the effect of dipole-dipole interactions. The immediate and surprising result to be drawn from the data presented is that charged grains show enhanced coagulation when compared to uncharged grains (figure 4). This behavior is primarily due to the fact that charge unequally distributed along the fractal surface results in a dipole contribution to the grain charge. In this case, colliding aggregates exert a torque on each other due to dipole-dipole interactions, altering the overall orientation of the aggregate as they stick. This allows smaller aggregates to coagulate more efficiently and produces, on average, denser particles with higher fractal dimension. However, aggregates growing from an initial population with a steep size distribution retain a broader distribution in fractal dimension (figure 9) and mean cluster mass (figure 10).

It is important to note that a quadratic growth rate for charged grains as seen in figure 5 indicates that while the overall growth rate is enhanced, runaway growth or charge gelation, characterized by $m \propto \exp(t)$, does not occur. Such runaway growth has been observed experimentally (Konopka *et al* 2005), and is surmised to be due to charge-dipole interactions between grains charged to opposite potentials due to triboelectric charging. This has also been confirmed through numerical simulations taking into account charge-dipole interactions between individual grains and the overall charge structure of the dust cloud (Matthews and Hyde 2004). Interestingly, these earlier simulations showed that like-charged grains exhibited slower growth when only charge-dipole interactions were allowed. In contrast, this work shows an increase in growth rate created in large part by dipole-dipole interactions and the effect that aggregate rotation has on the subsequent orientation of the colliding grains.

Although the overall distributions in mass and fractal dimension (figures 9 and 10) are strongly affected by the initial population distribution, the mechanism by which aggregation occurs appears to be independent, at least to some degree, of the exact size of the aggregates. In this study, both populations started with the same initial power-law distribution in aggregate radius. The minimum radii differed by a factor of two due to the difference in initial monomer size, while the maximum radii of the largest initial aggregates in the two populations were similar (cf. Table 1). Aggregation occurs in both populations of charged grains through the larger and mid-sized aggregates preferentially incorporating the smallest aggregates as evidenced in figures 7 and 8. Since this has not been seen in previous simulations of charged grains when all grains were negatively charged, even for situations where charge-dipole interactions were taken into account (Matthews and Hyde 2004), it is reasonable to assume this behavior to be due to the dipole-dipole interactions (i.e. the induced torques) between charged grains.

The fractal dimension is an important characteristic of an aggregate because of its influence on collision probabilities. Open structures with low fractal dimensions have a greater collisional cross-section, but they will also couple to the gas in their environment more efficiently, a process characterized by the gas friction time. Grains with similar gas friction times will tend to have small relative velocities and a smaller collision probability. The sharply peaked probability distribution in fractal dimension for the charged aggregates in Population II (figure 9c) presents an interesting case. It appears that fractal aggregates formed of monomers of similar size, with a narrow initial size range as in Population II, form clusters having similar fractal dimensions. This would indicate that such clusters develop similar gas-friction times as coagulation progresses, resulting in small relative velocities and preventing further collisions and growth. More realistic astrophysical dust populations, as represented by Population I, produce a more broadly peaked probability density distribution (figure 9a). This would indicate that grain growth in these populations could persist over greater time intervals given sufficient differences in the relative velocities between the grains driving coagulation. However, it is likely that a simulation which includes a larger number of particles would also lead to aggregates having similar sizes and friction times, narrowing the peak in the distribution function for fractal dimension as seen in Population II. Further studies are ongoing to examine the probable size limits for given environments including specific parameters for gas density, turbulence, and initial grain size.

Finally, the relatively small growth of the largest aggregates in the charged populations (figure 6) suggests that for coagulation to occur between like-charged grains, dust grains must have high relative velocities and/or relatively high densities in order to overcome the coulomb potential barrier and stick. Neutral grains may or may not produce a single large aggregate due to the fact that collisions between aggregates are purely random, having no driving force to select a certain growth pattern. Although the inclusion of dipole-dipole interactions can enhance growth rates for like-charged grains, the overall coagulation process itself tends to create grains of similar size. This in turn reduces the relative velocities between grains and may well provide a stopping mechanism for coagulation at some point. Thus at some point charged grains grow large enough to cut off coagulation unless some other mechanism exists which can continue to supply smaller grains. Since relative velocities between grains scale as the Keplerian velocity (Dominik *et al* 2007) and dust densities are generally greatest at the inner edge of a protoplanetary disk (Cuzzi 1993), the conditions most favorable for the growth of the largest charged aggregates are likely to exist within the inner regions of the protoplanetary disk. This has profound implications for current planet formation theories. An investigation of specific parameters in protoplanetary disks (including temperature, neutral gas, plasma, and dust densities, initial grain sizes and materials) will be the subject of a future investigation.

These results indicate that the microphysics of coagulation is quite sensitive to the initial conditions available, and not just on the gas to dust density ratio. The charge and charge arrangement on the forming aggregates is a significant factor in both the coagulation rate and the geometry of the resulting grains. Since the charging depends on many factors including plasma environment, ambient temperature, UV radiation, grain material, and transient heating events (Horányi and Goertz 1990), the amount of growth through coagulation will be unique to each specific environment. One of the outstanding problems in early planet formation is understanding the microphysics governing the collisions of dust particles in a protoplanetary disk, including realistic treatments of fractal dimension, porosity, collision outcomes, and charging, among other effects (Blum and Wurm 2008). Coagulation of charged aggregates also plays a role in atmospheric processes, such as the growth of tholin particles in the atmosphere of Titan (Rodin *et al* 2008). Extension of this work will allow the evolution of dust in such environments to be modeled in a self-consistent manner.

**Acknowledgements**
The authors wish to thank Victor Land for helpful discussions of the work.

**References**
Allen J E 1992 Probe theory - the orbital motion approach *Phys. Scr.* **45** 497-503
Blum J and G Wurm 2008 The growth mechanisms of macroscopic bodies in protoplanetary disks *Annu. Rev. Astro. Astrophys.* **46** 21-56
Cuzzi J N 1993 particle-gas dynamics in the mid-plane of a protoplanetary nebula, *Icarus* **106** 102-34
Dominik C, and A Tielens 1997 The physics of dust coagulation and the structure of dust aggregates in space *Ap. J.* **480** 647-73

Dominik C, J Blum, J N Cuzzi and G Wurm, 2007 Growth of dust as the initial step toward planet formation, *Protostars and Planets V*, eds B Reipurth, D Jewitt, K Keil, (Tucson: U. of Ariz. Press) pp 783-800

Goertz C K and G Morfill 1983 A model for the formation of spokes in Saturn's rings, *Icarus* **53**, 219-29

Horányi M 1996 Charged dust dynamics in the solar system *Ann. Rev. Astron. Asrophys.* **34** 383-413

Horányi M and C K Goertz 1990 Coagulation of dust particles in a plasma *Ap. J.* **361** 155-61

Ivlev A V, G M Morfill, and U Konopka 2002 Coagulation of charged microparticles in neutral gas and charge-induced gel transitions *Phys. Rev. Lett.* **89** 195502

Kempf S, S Pfalzner, T K Henning 1999 N-particle simulations of dust growth – I. Growth driven by Brownian motion *Icarus* **141** 388-98

Kennedy R V and J E Allen 2003 The floating potential of spherical probes and dust grains. II: Orbital motion theory *J. of Plasma Phys.* **69** 485-506

Konopka, U, F Mokler, A V Ivlev, et al 2005 Charge-induced gelation of microparticles *NJP* **7** 227

Laframboise J G and L W Parker 1973 Probe design for orbit-limited current collection *Physics of Fluids* **16** 629-36

Lampe M, G Joyce, G Ganguli and V Gavrishchaka 2001 Analytic and simulation studies of dust grain interaction and structuring *Phys. Scr.* **T89** 106-11

Lapenta G 1998 Effect of dipole moments on the coagulation of dust particles immersed in plasmas *Phys. Scr.* **57** 476-80

Matthews L S and T W Hyde 2004 Effects of the charge-dipole interaction on the coagulation of fractal aggregates *IEEE Trans. Plasma Sci.* **32** 586-93

Matthews, L S, R L Hayes, M S Freed, and T W Hyde 2007 Formation of cosmic dust bunnies, *IEEE Trans. Plasma Sci.* **35** 260-5

Matthews L and T W Hyde 2008 Charging and Growth of Fractal Dust Grains, *IEEE Trans. Plasma Sci,* **36** 310-4

Meakin P 1984 Fractal aggregates in geophysics *Rev. Geophys.* **29** 317

Richardson D C 1995 A self-consistent treatment of fractal aggregate dynamics *Icarus* **115** 320-35

Rodin A V, H U Keller, Y Skorov, et al,2008 Microphysical modeling of tholin haze in the Titan atmosphere, 37th COSPAR Scientific Assembly

Selwyn G, J E Heidenreich, and K Haller 1990 Particle trapping phenomena in radio frequency plasmas *Appl. Phys. Lett.* **57** 1867

Simpson I C 1978 The role of induced charges in the accretion of charged dust grains *Astrophys. and Sp. Sci.* **57** 381-400

Simpson I C, S Simons, and I P Williams 1979 Thermal coagulation of charged grains in dense clouds *Astrophys. and Sp. Sci.* **61** 65-80

Van Dongen P G J and M H Ernst 1985 Dynamic scaling in the kinetics of clustering *Phys. Rev. Lett.* **54** 1396-99

Weidenschilling S J and T V Ruzmaikina 1994 Coagulation of grains in static and collapsing protostellar clouds *ApJ* **430** 713-26

Winter J 1998 Dust in fusion devices – experimental evidence, possible sources and consequences *Plasma Phys. Control. Fusion* **40** 1201-10